\documentclass[11pt,twocolumn,tighten,twocolappendix]{aastex63}
\usepackage{hyperref}
\usepackage{color}
\usepackage{url}
\usepackage{natbib}
\usepackage{graphicx}
\usepackage{xfrac}
\usepackage{ulem}
\usepackage{CJK}
\usepackage{threeparttable}
\usepackage{multirow}
\usepackage{amsmath}
\begin{document}
\shorttitle{Magnetic fields in the CGM}
 \shortauthors{T.W. Lan \& J. Xavier Prochaska}
\title{Constraining Magnetic Fields in the Circumgalactic Medium} 
\begin{CJK*}{UTF8}{bsmi}
\author{Ting-Wen Lan 藍鼎文}
\affil{Department of Astronomy and Astrophysics, UCO/Lick Observatory, University of California, \\ 1156 High Street, Santa Cruz, CA 95064, USA}
\affil{Kavli IPMU, the University of Tokyo (WPI), Kashiwa 277-8583, Japan}
\author{J. Xavier Prochaska}
\affil{Department of Astronomy and Astrophysics, UCO/Lick Observatory, University of California, \\ 1156 High Street, Santa Cruz, CA 95064, USA}
\affil{Kavli IPMU, the University of Tokyo (WPI), Kashiwa 277-8583, Japan}

\begin{abstract}

We study the properties of magnetic fields in the circumgalactic medium (CGM) of $z<1$ galaxies by correlating Faraday rotation measures of $\sim 1,000$ high-redshift radio sources with the foreground galaxy number density estimated from  the DESI Legacy Imaging Surveys. This method enables us to extract  signals of rotation measures contributed by intervening gas around multiple galaxies. Our results show that there is no detectable correlation between the distribution of rotation measures and the number of foreground galaxies, contrary to several previous results. Utilizing the non-detection signals, we estimate $3\sigma$ upper limits to the rotation measures from the CGM of $\sim 20 \rm \  rad/m^{2}$ within~50 kpc and $\sim 10 \rm \ rad / m^{2}$ at separations of $100$ kpc. By adopting a column density distribution of ionized gas obtained from absorption line measurements, we further estimate the strengths of coherent magnetic fields parallel to the line of sight of $<\rm 2 \ \mu G$ in the CGM. 
We show that the estimated upper limits of rotation measures and magnetic field strengths are sufficient to constrain outputs of recent galaxy magneto-hydrodynamic simulations. Finally, we discuss possible causes for the inconsistency between our results and previous works.
\end{abstract}

\section{Introduction}
The circumgalactic medium (CGM) plays an important role in galaxy formation and evolution. 
It is the interface between the intergalactic medium and galaxies, where galaxies acquire gas fuel to supply star-formation and deposit metal-rich gas produced by feedback mechanisms driven by the explosion of stars and/or activities of supermassive black holes \citep[see][for a review]{Tumlinson2017}. 
In the past two decades, substantive progress has been made to better understand the properties of the CGM, including the gas content \citep[e.g.,][]{Stocke2013, Werk2014, Prochaska2017, Bordoloi2018}, the connections between the properties of galaxies and their CGM \citep[e.g.,][]{Chen2010, Kac2012, Bordoloi2011, Zhu2013b, Nielsen2013, Lan2014, Borthakur2016, Heckman2017, Lan2018, Schroetter2019}, and the small-scale structure of gas clouds \citep[e.g.,][]{Lan2017, Rubin2018, Peroux2018}. To interpret these observational results, theoretical and numerical studies have explored  various mechanisms that account for the observed properties of the CGM \citep[e.g.,][]{Nelson2018, Hummels2019, Voort2019, LanMo2019}. 

One of the potentially key physical components in the CGM is magnetic fields. Theoretically, the presence of magnetic fields has been found to prolong the lifetime of gas clouds in the CGM and also affect the morphology of gas structure  \citep[e.g.,][]{Chandran1998, McCourt2015, Ji2018, Berlock2019, Liang2020, Nelson2020}. Magnetic fields are also tightly connected to the properties of cosmic rays, another potentially key physical component of the CGM \citep[e.g.,][]{Salem2016,Hopkins2020}, and sensitive to the feedback mechanisms adopted in simulations \citep[e.g.,][]{Pakmor2013, Pakmor2019}.
However, despite the growing importance of magnetic fields in the CGM suggested by these theoretical and numerical works, observational constraints on the circumgalactic magnetic fields are still limited and uncertain.

Faraday rotation has been used as a powerful tool to probe magnetic fields in the Universe \citep[see][for a review]{Han2017}. When linearly polarized waves propagate through a magnetized plasma, the plasma will cause rotation of the polarization angle $(\rm \Delta \Phi)$ of the linearly polarized waves \citep[e.g.,][]{Klein2015}, 
\begin{equation}
    \rm \Delta \Phi = RM \, \lambda^{2}, 
\end{equation}
where $\lambda$ is the observed wavelength and the rotation measure, RM, is the integration of the product of electron density and magnetic field along the line of sight, 
\begin{equation}
    \rm RM \, [rad/m^{2}]= 0.812\int_{z_{s}}^{us} \frac{n_{e}}{(1+z)^{2}} \ B_{||} \cdot dl, 
    \label{eq:RM}
\end{equation}
where $n_{e}$ is electron number density ($\rm cm^{-3}$), $B_{||}$ is the magnetic field strength parallel to the sightline ($\mu G$), dl is the path length (parsec). 

By making use of rotation measures of high redshift radio sources, many studies have searched for possible signals of rotation measures induced
by intervening circumgalactic (and interstellar) medium
gas tracers, such as MgII absorbers \citep[e.g.,][]{Kronberg1982, Welter1984, Kronberg2008, Bernet2008, Bernet2010, Joshi2013, Farnes2014,  Malik2020} and Lyman alpha absorption systems \citep[e.g.,][]{Wolfe1992, Oren1995, Farnes2017}. However, these studies have not yielded a coherent picture of rotation measures from the CGM of galaxies.
The main limitation of such an analysis is that 
the rotation measures of those sightlines are mostly enhanced by the CGM of a single galaxy \citep[e.g.,][]{Lan2019}, whose signal could be too small to be robustly detected. In addition, absorption line systems are usually detected at redshifts greater than 0.4 in optical wavelengths, where the 
$(1+z)^{-2}$ factor in Equation~\ref{eq:RM} reduces the observed signal.
\cite{Prochaska2019} report the analysis of a fast radio burst (FRB)
whose sightline penetrates a massive foreground galaxy halo to 
constrain its circumgalactic magnetic field.
While this technique offers promise, it awaits the discovery 
of 100+ well-localized FRBs for a statistically powerful sample \citep[e.g.,][]{Ravi2019} as well as a better measurement of the Galactic foreground.

To overcome previous limitations, in this work we probe the rotation measures of the CGM of galaxies by correlating the rotation measures observed for
high-redshift radio sources with the number of photometric galaxies in the foreground. This analysis makes use of the fact 
that the ionized CGM is ubiquitously detected 
around $10^{8}-10^{11} \ M_{\odot}$ galaxies 
\citep[with almost $100\%$ covering fraction within 200~kpc; e.g.,][]{Stocke2013, Werk2014, Prochaska2017, Bordoloi2018}. 
In so doing, we utilize information on the full foreground galaxy density fields to extract the rotation measure signals introduced by the CGM of foreground galaxies. 
The structure of the paper is as follows. Our data analysis is described in Section 2. We show our results and discuss their implications in Section 3. We summarize in Section 4. Throughout the paper we adopt a flat $\Lambda$CDM cosmology with $h=0.7$ and 
$\Omega_{\rm M}=0.3$.

\section{Data analysis}
\subsection{Method}
The observed rotation measure of an extragalactic radio source has contributions from the Milky Way, the source, and any gaseous structure in between.  This can be expressed as 
\begin{equation}
    \rm RM_{obs} = RM_{Milky Way} + RM_{intervening} + RM_{source}.
\end{equation}
The intervening contribution can be further expressed as 
\begin{equation}
    \rm RM_{intervening} = \sum_{i}^{N_{gas}} RM_{gas}^{i}
\end{equation}
according to the number of gas structures, $\rm N_{gas}$, intercepted by the sightline. 
In this work, we use the number of foreground galaxies, $\rm N_{gal}$,  with impact parameters within 200 kpc as a proxy for the number of gas structures based on the fact that the CGM has nearly $100\%$ covering fraction at this impact parameter \citep[e.g.,][]{Bordoloi2018}. In what follows, we replace $\rm N_{gas}$ with $\rm N_{gal}$.

The rotation measure value can be positive and negative depending on the intercepting magnetic field orientation. If the coherent length of magnetic fields is much smaller than the path length of the sightline, the $\rm RM_{intervening}$ distribution is expected to follow a random walk process \citep[e.g.,][]{Akahori2010} and for sightlines intercepting $\rm N_{gal}$ gas structures,
the mean $\rm RM_{intervening}$ is expected to be consistent with zero and the standard deviation, $\rm \sigma(RM_{intervening})$, to be proportional to $\sqrt{N_{gal}}$. In other words, by investigating $\rm \sigma(RM_{intervening})$ as a function of $\rm N_{gal}$, one can extract the characteristic rotation measure contributed by intervening gas structure.

To extract the intervening RM signals, we obtain residual rotation measures, RRMs, where
\begin{equation}
    \rm RRM_{obs} = RM_{obs} - GRM,
\end{equation}
with GRM, the Galactic component obtained from a Galactic rotation measure map from \citet{Oppermann2015}. The map is reconstructed by an algorithm, based on information field theory \citep[e.g.,][]{Ensslin2009}, applied to $\sim 40,000$ rotation measures of sources on the sky.
We then measure the standard deviation of RRMs, $\rm \sigma_{RRM}$ as a function of number of foreground galaxies. We select radio background sources at high redshifts to suppress the source contribution, $\rm RM_{source}$. With the datasets described in the following section, we are able to probe sightlines where the number of galaxies ranges from 0 to 10, much wider than the range probed by previous studies using absorption line systems (mostly 0 or 1).

\subsection{Datasets}
\subsubsection{Galaxy catalog} 
To explore the correlation between rotation measures and the number of foreground galaxies, we make use of the galaxy catalog provided by the DESI Legacy Imaging Surveys \citep{Dey2019}, which consists of three public surveys, the Dark Energy Camera Legacy Survey, the Beijing-Arizona Sky Survey \citep{Zou2017}, and the Mayall z-band Legacy Survey. These surveys together cover about 14,000 deg$^{2}$ of the sky with $g$, $r$, and $z$ bands with 24, 23.4, and 22.5 limiting magnitudes for each band respectively. These depths are sufficient to detect $L*$ galaxies at redshift 1. We use the latest version of the galaxy catalog from the 8th data release\footnote{\href{http://legacysurvey.org/dr8/description/}{legacysurvey.org/dr8/description/}} with galaxy properties characterized by the {\it Tractor} algorithm\footnote{\href{https://github.com/dstndstn/tractor}{github.com/dstndstn/tractor}} \citep{Lang2016} \citep[See also Section 8 in][]{Dey2019}. The catalog consists of nearly 1~billion galaxies.

\begin{figure}
\center
\includegraphics[width=0.49\textwidth]{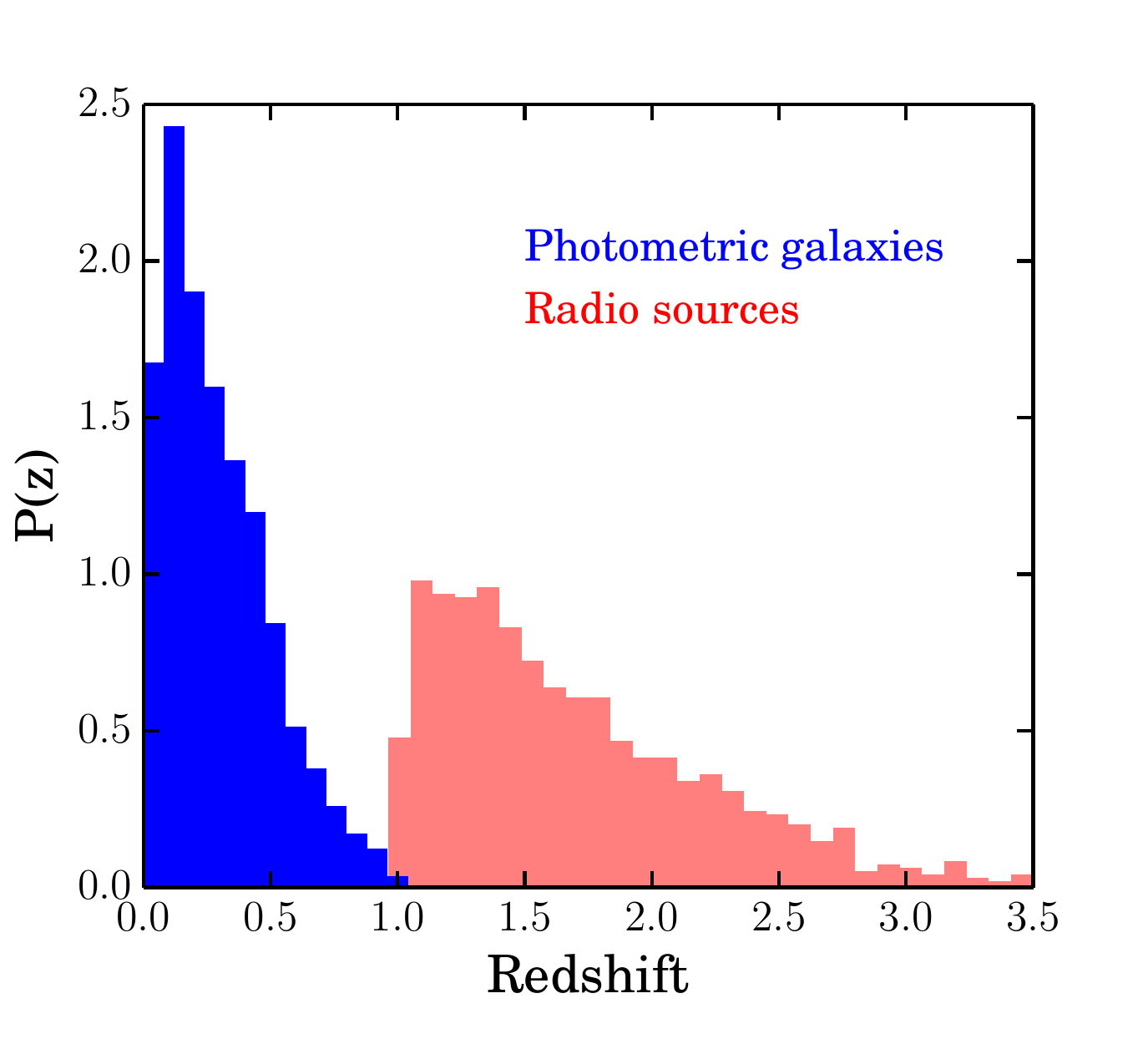}
\caption{Redshift distributions of the analyzed galaxy sample (blue) drawn from \cite{Zhou2020} and the radio background sources (red) from \citet{Farnes2014}.}

\label{fig:photoz}
\end{figure}

In this analysis,  we select galaxies with $z$-band magnitudes between 16 and 21 and photometric redshifts smaller than 1 by making use of the photometric redshifts provided by \citet{Zhou2020} based on a random forest algorithm. Figure~\ref{fig:photoz} shows the redshift distribution of the galaxy sample with the mean redshift $z \sim 0.3$. 

Finally, for each galaxy from the Legacy Surveys in our analysis, 
we estimate its stellar mass via spectral energy distribution fitting with  the CIGALE package \citep{Boquien2019}. We adopt the simple stellar population from \citet{BC2003} with 
Chabrier initial mass function \citep{Chabrier2003} and delayed-exponential star-formation history. The best-fit stellar mass is constrained by the observed flux in $g$, $r$, $z$, WISE~1 and WISE~2 bands. We provide the galaxy information at \url{https://people.ucsc.edu/~tlan3/research/CGM_magnetic_fields/}.

\begin{figure}
\center
\includegraphics[width=0.47\textwidth]{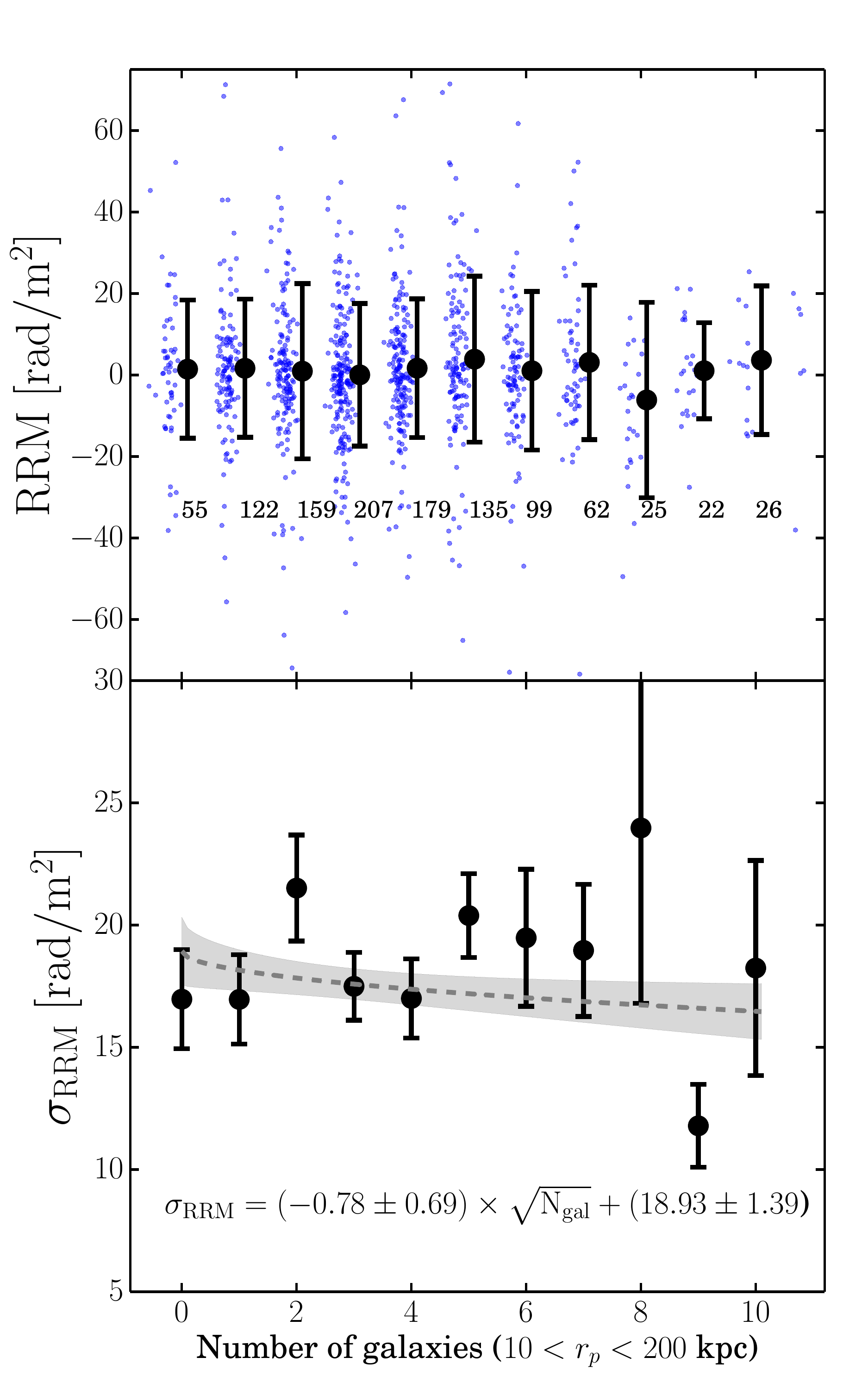}
\caption{\emph{Top:} Residual rotation measures (RRMs) as a function of number of galaxies 
$N_{\rm gal}$ within estimated physical impact parameter $10<r_{p}<200$ kpc. The blue data points show the individual sightlines. The black data points show the mean values of RRMs and the errorbars show the corresponding dispersion of the individual data points, estimated from the standard deviation. The number of sightlines for each is listed under the black data points. 
\emph{Bottom:} Dispersion of RRMs, $\sigma_{\rm RRM}$, as a function of number of galaxies. The errorbars show the bootstrapping uncertainties for $\sigma_{\rm RRM}$. The grey dashed line and shaded region show the best-fit function (Equation~\ref{eq:ab}) and the corresponding $1\sigma$ region.  
}
\label{fig:sigma_angular}
\end{figure}

\subsubsection{Rotation measure catalog} 
We use rotation measures of extragalactic sources from \citet{Farnes2014} as our primary dataset. This catalog is mainly based on the rotation measures of $\sim 35,000$ sources from \citet{Taylor2009}.  
In addition, it includes the redshift information of the extragalactic sources compiled by \citet{Hammond2012}. 

Given that our main goal is to extract the rotation measures contributed by the CGM of intervening galaxies, we only select radio
sources with redshift $z>1$ as background. The redshift distribution of the radio background sources is shown in Figure~\ref{fig:photoz}. This selection also reduces the contribution of rotation measures intrinsic to the background objects
(RM$_{\rm source}$). Moreover, this selection is motivated by the fact that galaxies brighter that 21 magnitude in $z$-band are mostly at redshift lower than 1 (Figure~\ref{fig:photoz}). Therefore, nearly all of 
the photometric galaxies detected close on the sky to
the background sources are expected to contribute to 
their rotation measure.

Finally, we only select sources within the footprint of the Legacy Surveys with at least three exposures in $z$-band and $g$-band and with the 5$\sigma$ limiting magnitude in $z$-band of extended sources being deeper than 22.5\,mag. To ensure that our measurements are not sensitive to a few outliers, we remove sightlines with $\rm |RRM|>100$ $\rm  rad/m^{2}$ and with the uncertainty of RRM greater than 20 $\rm rad/m^{2}$. 
This selection removes $3\%$ of the sources.
The final catalog consists of $\sim 1,100$ background sources. 
 
\subsubsection{Galactic map of rotation measures} 
To remove the Galactic component, we adopt the map of GRMs provided by \citet{Oppermann2015}\footnote{\url{https://wwwmpa.mpa-garching.mpg.de/ift/faraday/2014/index.html}}. These authors develop a reconstruction algorithm within the framework of information field theory \citep[e.g.,][]{Oppermann2011, Oppermann2012} and apply it to the rotation measures of $\sim 40,000$ radio sources on the sky compiled from the literature. The map has $\sim 30'$  angular resolution. For each sightline of our sample, we use the GRM value of the nearest bin from the map and subtract it from the rotation measure of the radio source. The uncertainty of the RRM of the sightline is estimated with the quadrature of the uncertainties of the RM and GRM. We note that \citet{Xu2014a} also produce a map of Galactic rotation measures, which has angular resolution $\sim 3$~degrees,  based on a weighted-average method. We have performed our analysis using the RRM data from \citet{Xu2014b} which is based on the Galactic map of \citet{Xu2014a} and found consistent results (see Appendix \ref{app:A}).

\section{Results and discussion}

In this section, we present results of our correlation measurements between the residual rotation measures of background objects and the number of photometric galaxies detected in the Legacy Surveys. We will explore how $\sigma_{\rm RRM}$ depends on the number of galaxies $N_{\rm gal}$
along the sightlines and use such quantities to constrain the rotation measures and corresponding magnetic field strengths of the gas around galaxies. 

\subsection{$\sigma_{\rm RRM}$ as a function of number of galaxies}

We first explore $\sigma_{\rm RRM}$ as a function of number of galaxies with impact parameters $r_p < 200$~kpc. The impact parameters are estimated based on the photometric redshifts of galaxies and angular separation between the background and the galaxies.
The results are shown in Figure~\ref{fig:sigma_angular}.
The top panel shows the distribution of RRM as a function of number of galaxies. 
The black data points show the average RRM and the errorbars indicate the standard deviation of RRM, $\sigma_{\rm RRM}$. As expected, the average RRM is consistent with zero and there is no correlation between the average RRM and the number of galaxies. 

The lower panel of Figure~\ref{fig:sigma_angular} shows the standard deviation of the residual rotation measures, $\sigma_{\rm RRM}$, as a function of $N_{\rm gal}$. The corresponding uncertainties are calculated by bootstrapping the catalog 500 times. Interestingly, we find no significant correlation between $\sigma_{\rm RRM}$ and the number of galaxies. 
To characterize the significance of the trend, we fit the measurements with 
\begin{equation}
    \sigma_{\rm RRM}=A\times \sqrt{N_{\rm gal}}+B,
    \label{eq:ab}
\end{equation}
and find the best fit A and B parameter values are $-0.8\pm0.7$ $\rm rad/m^{2}$ and $18.9\pm 1.4$ $\rm rad/m^{2}$ respectively. This is consistent with no correlation between $\sigma_{\rm RRM}$ and the number of galaxies.

\begin{figure}
\center
\includegraphics[width=0.45\textwidth]{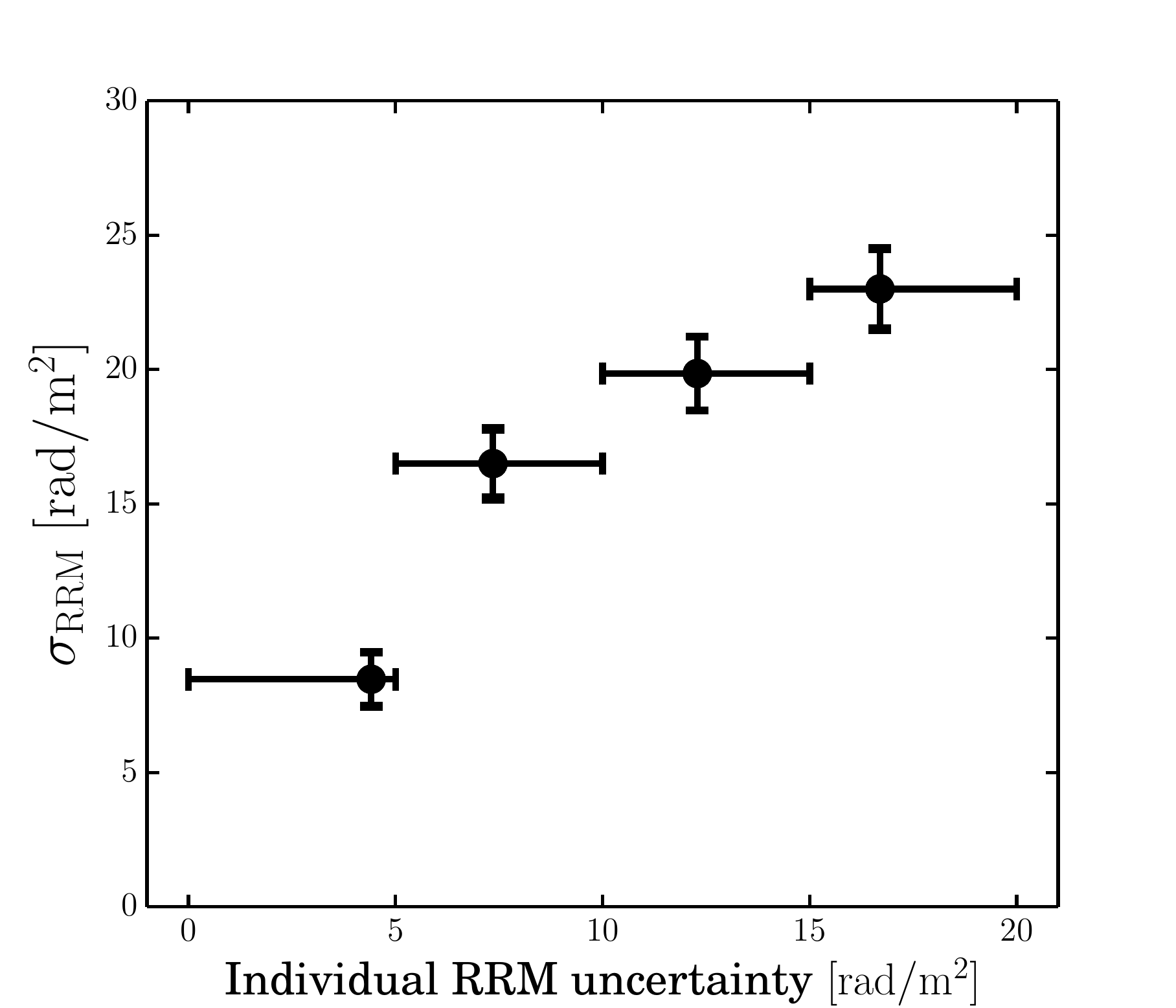}

\caption{Dispersion of RRMs as a function of measured uncertainties of individual sightlines.  The tight correlation between $\sigma_{\rm RRM}$ and the RRM uncertainty indicates that the latter dominates the former.
}
\label{fig:sigma_uncertainty}
\end{figure}

\begin{figure*}
\center
\includegraphics[width=1\textwidth]{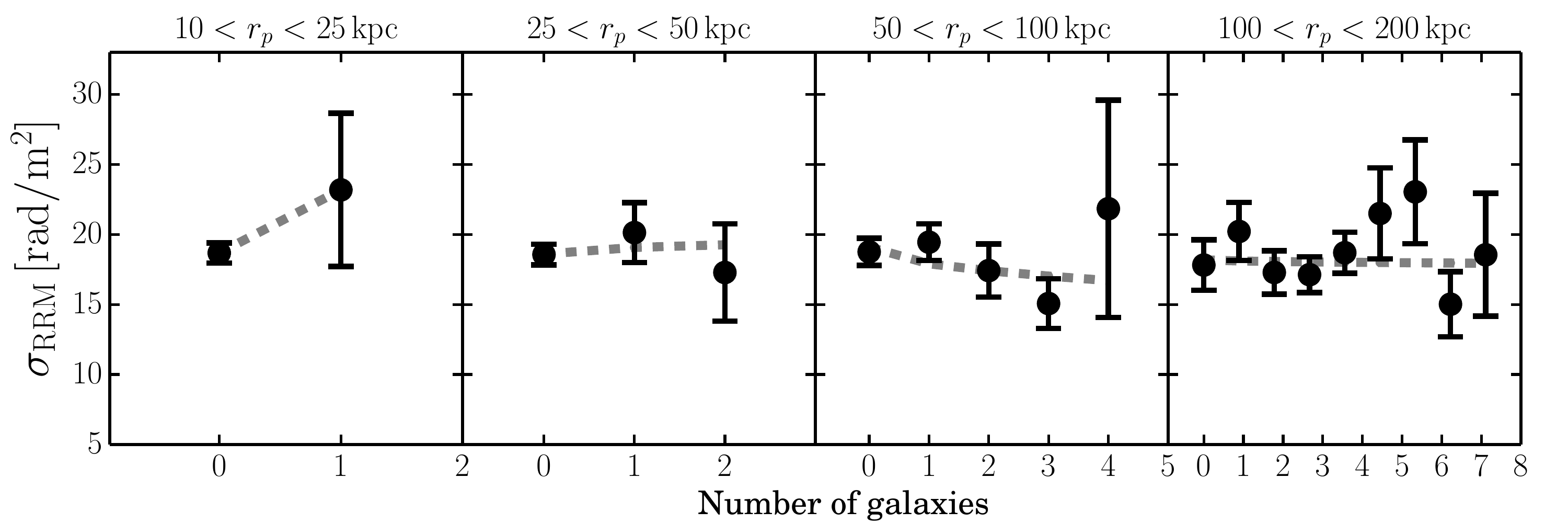}
\caption{Dispersion of RRMs as a function of number of galaxies and impact parameters from 10 kpc to 200 kpc. The gray dashed lines show the best-fit functions. No correlation between $\sigma_{\rm RRM}$ and $\rm N_{gal}$ is detected at all scales. 
}
\label{fig:sigma_rrm_kpc}
\end{figure*}
\begin{figure}
\center
\includegraphics[width=0.45\textwidth]{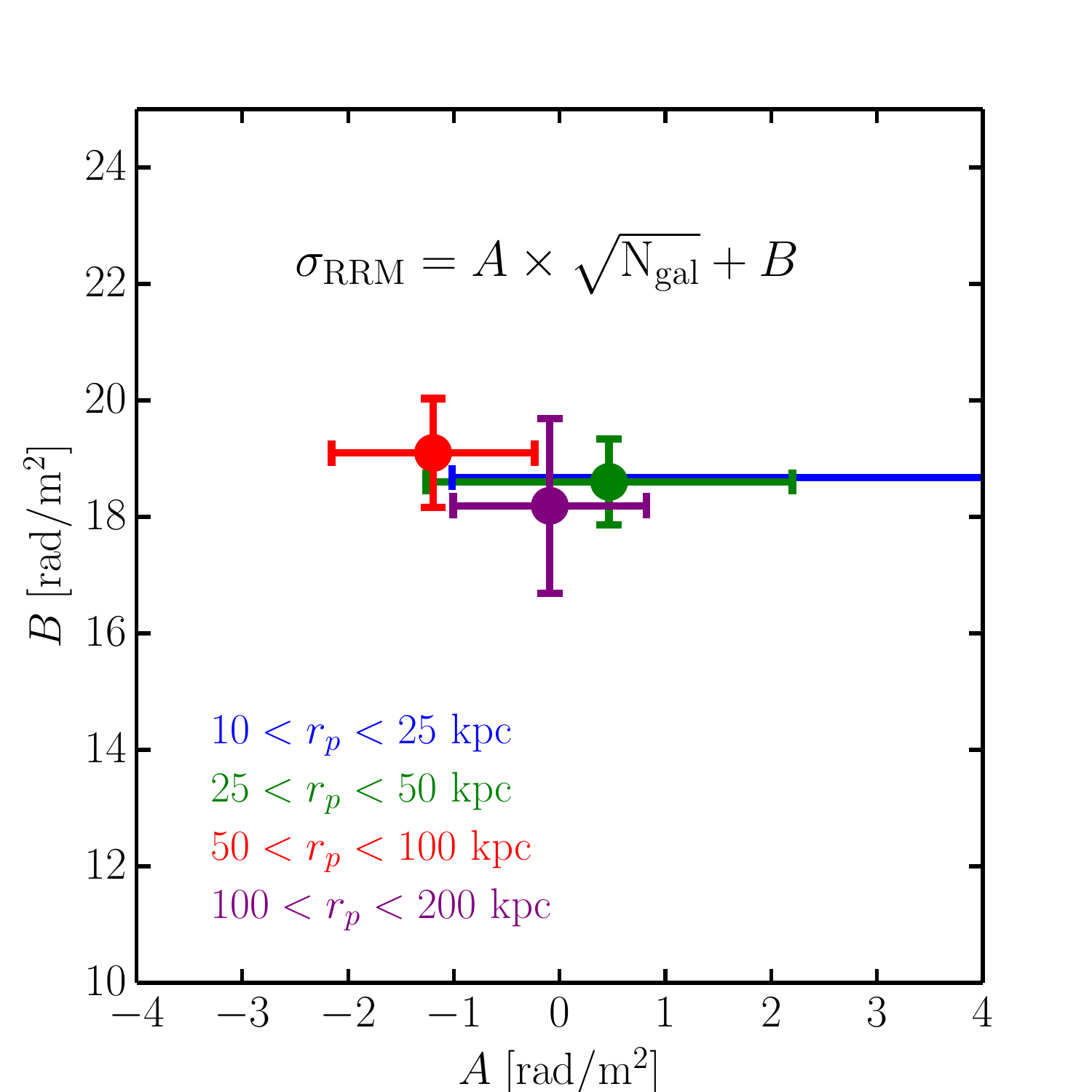}
\caption{Best-fit parameter A \& B values as a function of impact parameters. Independent of the $r_p$ interval, the A values are consistent with zero and the B values are consistent with 18. 
}
\label{fig:para}
\end{figure}
\begin{figure*}
\center
\includegraphics[width=0.95\textwidth]{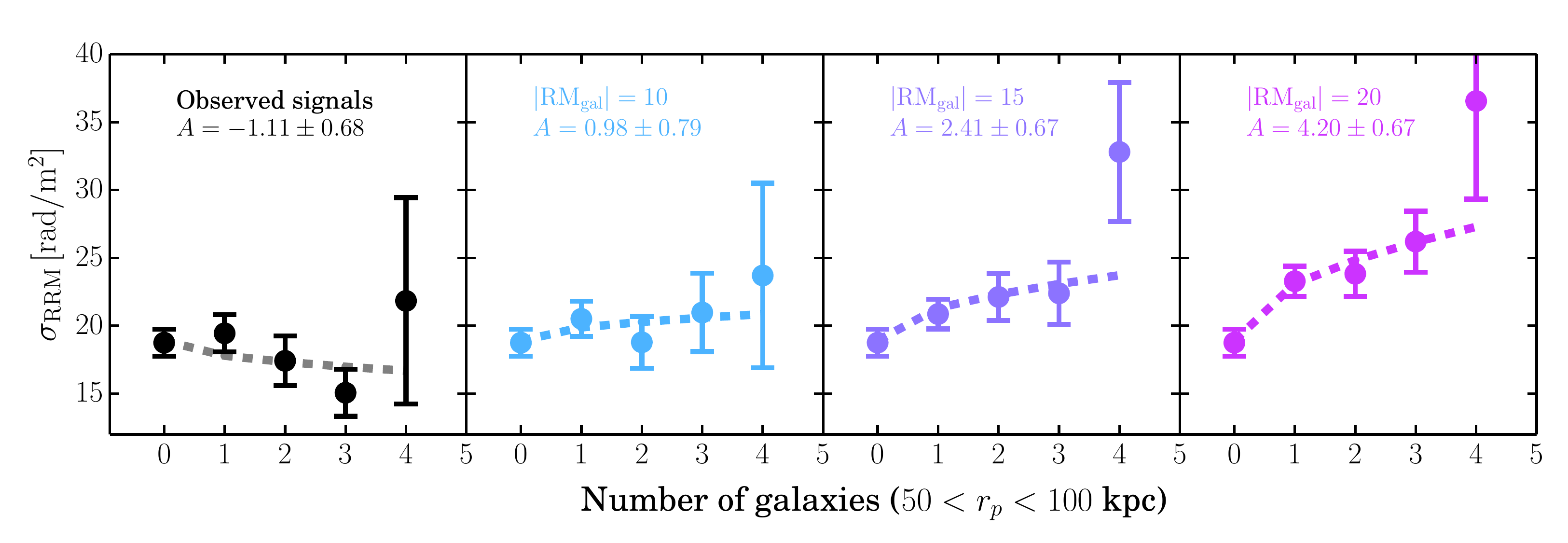}

\caption{Illustration of the process for estimating the upper limits of the rotation measure contribution from galaxies. The first panel from the left shows the observed $\sigma_{\rm RRM}$ as a function of number of galaxies from $50<r_{p}<100$ kpc bin. The second to the fourth panels show the expected $\sigma_{\rm RRM}$ as a function of $\rm N_{gal}$ if each galaxy contributes $\rm |RM_{gal}|=10, 15, and \ 20 \ rad/m^{2}$ respectively. We consider a $|RM_{gal}|$ value as $3\sigma$ upper limit when A parameter is detected with $3\sigma$ significance. 
}
\label{fig:illustration_simulation}
\end{figure*}

We further note that $\sigma_{\rm RRM}$ is $\approx 20 \ \rm rad/m^{2}$ for each of the bins. This zero point primarily reflects the uncertainties of the residual rotation measures. To confirm this assertion, we select sightlines with different RRM uncertainties and calculate $\sigma_{\rm RRM}$ as a function of RRM uncertainty. The result is shown in Figure~\ref{fig:sigma_uncertainty}. As can be seen, $\sigma_{\rm RRM}$ increases with the RRM uncertainty, demonstrating that the zero point seen in the bottom panel of Figure~\ref{fig:sigma_angular} is dominated by the measurement uncertainties. 

The non-detection of a correlation between $\sigma_{RRM}$ and $N_{\rm gal}$, shown in Figure~\ref{fig:sigma_angular}, demonstrates that with the precision of the rotation measures from the dataset, the contribution of rotation measures from intervening galaxies is too small to be detected.  Nevertheless, we can utilize the non-detection to place constraints on the rotation measures from the CGM of galaxies. 

Finally, we note that this non-detection result is inconsistent with several previous results, which report a significant detection between the presence of MgII absorbers and excess rotation measures \citep[e.g.,][]{Bernet2008, Farnes2014, Malik2020}. We have investigated possible causes for the inconsistency and found that the galactic map used in those analysis may affect the results. The details of the investigation are described in Appendix~\ref{app:B}.

\subsection{Constraining the rotation measures around galaxies}
Before constraining the rotation measures around galaxies, we first measure $\sigma_{\rm RRM}$ as a function of number of galaxies and impact parameters.  Figure~\ref{fig:sigma_rrm_kpc} shows the results in bins of
impact parameter from 10 kpc to 200 kpc and Figure~\ref{fig:para} shows the corresponding best fit parameter values for A and B from Equation~\ref{eq:ab}. Similarly to Figure~\ref{fig:sigma_angular}, we find no correlation between $\sigma_{\rm RRM}$ and the number of galaxies from 10 to 200 kpc.

We now constrain the contribution of the rotation measures from the CGM of galaxies. To do so, we perform simulations by adding rotation measures to the original RRM values and perform the measurements iteratively until the best-fit A parameter value in Equation~\ref{eq:ab} exceeds 0 with $3\sigma$ significance. In this way, the uncertainties and the underlying distribution of RRM of the dataset are preserved in the estimation. 

More specifically, we first assume an intrinsic rotation measure from a galaxy, $\rm |RM_{gal}|$, and for each sightline, we add the contribution to the original RRM value as 
\begin{equation}
    \rm RRM_{simulated}^{i} = RRM_{original}^{i}+\sum_{j}^{N_{gal}} \pm \frac{|RM_{gal}|}{(1+z_{gal,j})^{2}}
\end{equation}
where $i$ indicates sightline $i$, $N_{gal}$ is number of galaxies for the sightline, $z_{gal,j}$ is the photo-z of galaxy $j$ around the sightline, and $\pm$ reflects the random walk nature of the rotation measure. After this process for all the sightlines, we calculate $\sigma_{\rm RRM}$ as a function of $N_{\rm gal}$ and obtain the best-fit A parameter value.
While performing the fitting process, we fix the B parameter value to be the global $\rm \sigma_{RRM}$ of the sample, 18.9 $\rm rad/m^{2}$. Figure~\ref{fig:illustration_simulation} illustrates the process. The left panel shows the original measurements for $50<r_{p}<100$ kpc bin as shown in Figure~\ref{fig:sigma_rrm_kpc}. From the second to the fourth panel, we show the $\sigma_{\rm RRM}$
measurements recovered after introducing $\rm |RM_{gal}|=10, 15, 20 \ rad/m^{2}$ for each galaxy respectively. As expected, the simulated $\sigma_{\rm RRM}$ increases with the number of galaxies, and the parameter A value increases with $\rm |RM_{gal}|$. The S/N of the A parameter reaches 3$\sigma$ with $\rm |RM_{gal}|\simeq 15 \ rad/m^{2}$; this value may be considered as the $3\sigma$ upper limit for the 
characteristic rotation measure of the CGM of galaxies. 

\begin{figure*}
\center
\includegraphics[width=1\textwidth]{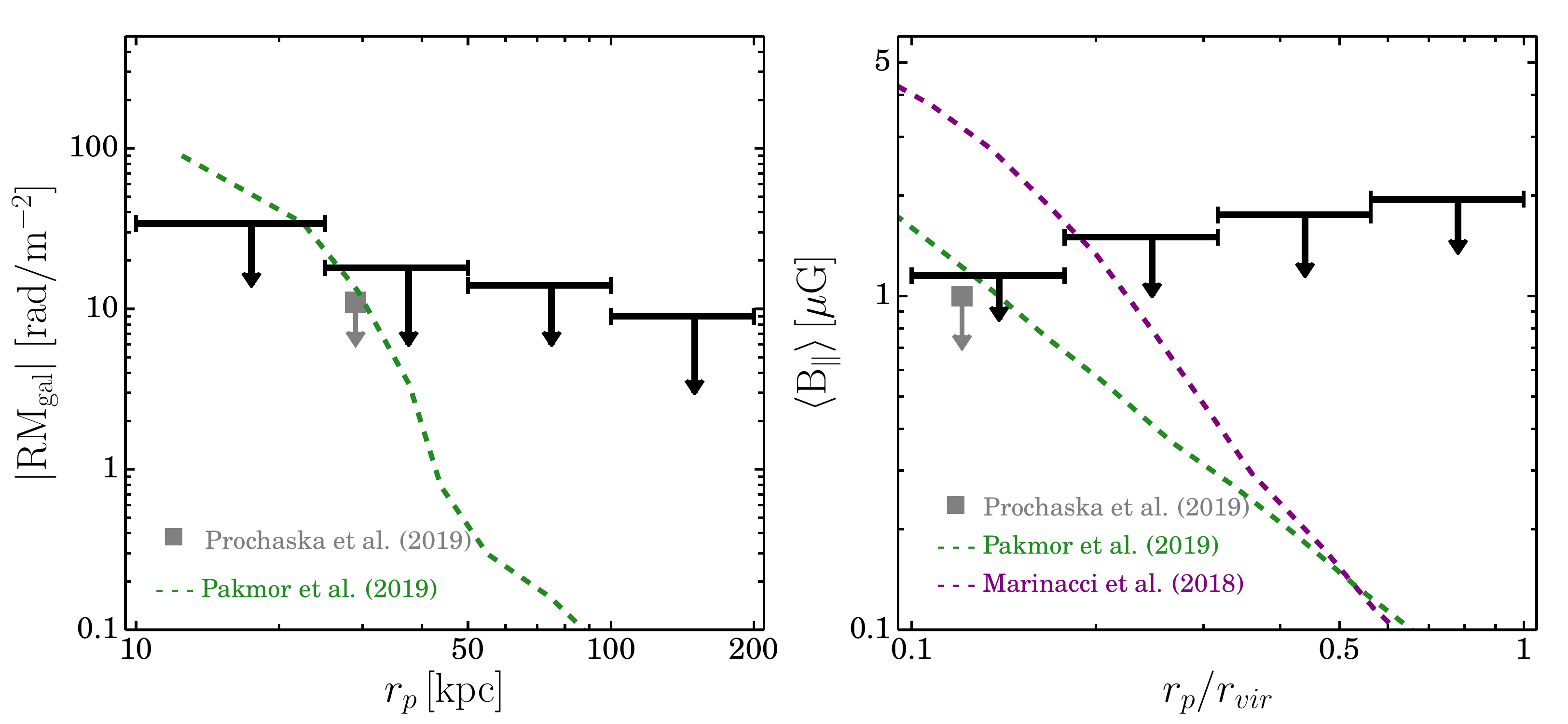}
\caption{\emph{Left}: $3\sigma$ upper limits of rotation measures from the CGM of galaxies as a function of impact parameter. \emph{Right:} $3\sigma$ upper limits of the strengths of coherent magnetic fields parallel to the line of sight as a function of $r_{p}/r_{vir}$. 
The grey data points show the constraints around a $M* \approx 10^{10.7}\ M_{\odot}$ galaxy from \citet{Prochaska2019}. The green and purple dashed lines show the simulated properties of Milky Way-type galaxies from magnetohydrodynamic simulations by \citet{Pakmor2019} and \citet{Marinacci2018}. We note that the green dashed line in the left panel is measured in projected distance, the same as the observations, while the lines in the right panel are measured in 3D instead of projected space. 
}
\label{fig:RM_limit}
\end{figure*}

We perform this analysis for all the impact parameter bins and the results are shown in the left panel of Figure~\ref{fig:RM_limit}. We find that the rotation measure from the CGM of galaxies is smaller than 30 $\rm rad/m^{2}$ within 30 kpc and smaller than 10 $\rm rad/m^{2}$ around 200 kpc. 
The tighter upper limit at higher $r_p$ reflects the wider range of $\rm N_{gal}$ probed by the background sources.
The green dashed line shows the rotation measure as a function of impact parameter from recent hydrodynamic simulations by \citet{Pakmor2019}. As can be seen, our upper limit within $30$ kpc indicates that the simulation predicts too high rotation measures. As \citet{Pakmor2019} reported that the rotation measures in their simulation are driven by galactic outflows, our upper limit measurements have placed constraints on the models of galactic outflows. The grey data point shows a similar upper limit from \citet{Prochaska2019} with the rotation measure of a fast ratio burst in the background of a massive galaxy with $10^{10.7}\ M_{\odot}$. We also note that our constraints are consistent with the results from \citet{Oppermann2015} and \citet{Schnitzeler2010}, suggesting that the observed RM contributed from extragalactic sources is most likely lower than 6-7 $\rm rad/m^{2}$, which corresponds to $\sim 10\rm \ rad/m^{2}$ at redshift 0.3, the mean redshift of our galaxy sample.

\subsection{Constraining the strength of magnetic fields around galaxies}
We now estimate the upper limits of the strength of coherent magnetic fields parallel to the line of sight. Similarly to the process in Section 3.2, we obtain the upper limits by simulating realizations with different rotation measures from galaxies. This time, instead of assigning the value of rotation measure directly, we estimate the rotation measure based on the strengths of magnetic fields $\langle B_{||}\rangle$ and an adopted column density density of electrons $N_{e}$ as
\begin{equation}
   \rm |RM_{\rm gal}|=\beta\, N_{e}\, \langle B_{||}\rangle
\end{equation}
where $\beta=2.63\times 10^{-19}\rm \, rad \, m^{-2} \, cm^{2}\,  \mu G^{-1}$, $N_{e}$ has units of $\rm cm^{-2}$ and $\langle B_{||}\rangle$ is in the units of $\mu G$ \citep[e.g.,][]{Kronberg1982}. 
For the electron column density, we adopt the best-fit profile
\begin{equation}
    N_{\rm H} = 10^{19.1}\times \bigg(\frac{r_{p}}{r_{vir}}\bigg)^{-1}\rm  [cm^{-2}]
\end{equation}
from \citet{Werk2014} based on photo-ionization modeling for all the COS-Halo absorption line measurements \citep[][]{Tumlinson2013}. We adopt $N_{e} = 1.2 N_{\rm H}$ given that the gas is mostly ionized with ionization fraction $>99\%$ and to account for doubly-ionized Helium \citep{Werk2014}. 
In addition, we estimate the virial radius of the dark matter halos with a scaling relation. 
\begin{equation}
    r_{vir} = \rm 200 \ kpc \ \bigg(\frac{M_{*}}{10^{10.3} M_{\odot}}\bigg)^{1/5},
\end{equation}
from \citet{Zhu2013b} and perform the measurements with respect to the virial radius. This takes into account the dependence of column density and magnetic field strengths on the halo mass of galaxies.

The right panel of Figure~\ref{fig:RM_limit} shows the $3\sigma$ upper limits for $\langle B_{||}\rangle$ within the virial radius. Our results show that $\langle B_{||}\rangle$ is smaller than 2 $\mu G$ within the viriral radius and smaller than 1 $\mu G$ within 0.2 $r_{vir}$, which is consistent with the constraint from \citet{Prochaska2019} indicated by the grey data point. We also show simulated magnetic field strengths from \citet{Marinacci2018} and \citet{Pakmor2019}, indicated by the purple and green dashed lines (for $\sim 10^{12} \, M_{\odot}$ halo) respectively. We note that the simulated results are average magnetic field strengths measured in 3D. Therefore, a direct comparison between our upper limits and the simulated predictions can not be made. 
Nevertheless, it illustrates that our estimated upper limits of magnetic field strengths within 0.2 $r_{vir}$ are comparable to simulated ones and could effectively constrain model predictions when comparing to the same 2D measurements from simulations. 

Our estimated upper limits are consistent with the expected magnetic field strengths between galaxies and the intergalactic medium.
The strengths of magnetic fields of Milky Way-type galaxies have been measured to be in the order of $10 \ \mu G$ \citep[e.g.,][]{Beck2015, Basu2013, Mao2017} and the strengths of magnetic fields in the intergalactic medium, are expected to be much weaker, on the order of $nG$ \citep[e.g.,][]{Fukugita2004, Ryu2008, Vazza2018, Vernstrom2019, Sullivan2020, Stuardi2020}. 
We also note that the magnetic field strengths in the intracluster medium are $\sim 1 -  4 \ \mu G$ \citep[e.g.,][]{Govoni2010} at the same order of our constraints. Finally, we emphasize that the estimated upper limits of magnetic field strengths depend on the assumed gas column density. In our estimation, we only consider cool ionized gas with temperature around $10^{4}$ K. If a hot gas component with $\rm 10^{6}\ K$ is also considered \citep[e.g.,][]{Li2018,Lim2020}, we expect that the values of the upper limits will be a factor of two lower than shown in Figure~\ref{fig:RM_limit} albeit subject to uncertainties in the hot gas profile.

\section{Summary}
By correlating the residual rotation measures of background radio sources with the number of photometric galaxies detected in the DESI Legacy Imaging Surveys, we investigate the rotation measures contributed by the CGM of intervening galaxies. Our results are summarized as follows:
\begin{enumerate}
    \item No correlation between $\sigma_{\rm RRM}$ and $N_{\rm gal}$ is detected within the impact parameters analyzed ($10 < r_p < 200$kpc),
    contrary to some previous results obtained by the correlation between MgII absorbers and rotation measures. 
    
    \item We estimate $3 \, \sigma$ upper limits of RMs contributed by the CGM as a function of impact parameter and find the characteristic
    CGM RM is smaller than $30 \rm \ rad/m^{2}$ within 30 kpc, a value lower than the output signal of recent magneto-hydro simulations by \citet{Pakmor2019}. 
    
    \item Adopting a column density of ionized gas obtained from absorption line measurements, we estimate the strengths of coherent magnetic fields parallel to the sightlines as lower than $2 \rm \ \mu G$ in the CGM.
    
    \item Finally, we revisit some of the previous results showing $>3 \sigma$ detections of correlations between the presence of MgII absorbers and the rotation measure distribution and find that some of the results are sensitive to the foreground galactic model used in the analyses. If adopting the same galactic map, the results obtained from datasets used previously become consistent with ours.   
\end{enumerate}
In our analysis, the uncertainties of the RMs ($\sim 10-20 \rm \, rad/m^{2}$) and the precision of the Galactic map are the main limitations for detecting the magnetic field signals of the CGM. However, these limitations will soon be overcome by upcoming datasets from new radio surveys, such as LOFAR \citep[][]{Haarlem2013}, POSSUM \citep[][]{POSSUM} and SKA \citep{Carilli2004}, which will provide high precision rotation measures with uncertainties $\sim 1 \rm \, rad/m^{2}$ as well as high number density of radio sources across the sky. The combination of these radio datasets with galaxy information from optical imaging and spectroscopic surveys, such as LSST \citep{Ivezic2019}, Euclid \citep{Amiaux2012}, and DESI \citep{Levi2013}, will largely improve the constraints of rotation measures and magnetic field strengths of the CGM.
With new measurements, we will have a better understanding of the properties of magnetic fields around galaxies and how magnetic fields regulate galaxy evolution.

\acknowledgements
We thank Masataka Fukugita, Bryan Gaensler, and Rongman Bordoloi for useful discussions. 
TWL and JXP acknowledge support from NSF grant AST-1911140.
Kavli IPMU is supported by World Premier International Research Center Initiative of the Ministry of Education, Japan.

The Legacy Surveys consist of three individual and complementary projects: the Dark Energy Camera Legacy Survey (DECaLS; NOAO Proposal ID $\#$ 2014B-0404; PIs: David Schlegel and Arjun Dey), the Beijing-Arizona Sky Survey (BASS; NOAO Proposal ID $\#$ 2015A-0801; PIs: Zhou Xu and Xiaohui Fan), and the Mayall z-band Legacy Survey (MzLS; NOAO Proposal ID $\#$ 2016A-0453; PI: Arjun Dey). DECaLS, BASS and MzLS together include data obtained, respectively, at the Blanco telescope, Cerro Tololo Inter-American Observatory, National Optical Astronomy Observatory (NOAO); the Bok telescope, Steward Observatory, University of Arizona; and the Mayall telescope, Kitt Peak National Observatory, NOAO. The Legacy Surveys project is honored to be permitted to conduct astronomical research on Iolkam Du'ag (Kitt Peak), a mountain with particular significance to the Tohono O'odham Nation.

NOAO is operated by the Association of Universities for Research in Astronomy (AURA) under a cooperative agreement with the National Science Foundation.

This project used data obtained with the Dark Energy Camera (DECam), which was constructed by the Dark Energy Survey (DES) collaboration. Funding for the DES Projects has been provided by the U.S. Department of Energy, the U.S. National Science Foundation, the Ministry of Science and Education of Spain, the Science and Technology Facilities Council of the United Kingdom, the Higher Education Funding Council for England, the National Center for Supercomputing Applications at the University of Illinois at Urbana-Champaign, the Kavli Institute of Cosmological Physics at the University of Chicago, Center for Cosmology and Astro-Particle Physics at the Ohio State University, the Mitchell Institute for Fundamental Physics and Astronomy at Texas A$\&$M University, Financiadora de Estudos e Projetos, Fundacao Carlos Chagas Filho de Amparo, Financiadora de Estudos e Projetos, Fundacao Carlos Chagas Filho de Amparo a Pesquisa do Estado do Rio de Janeiro, Conselho Nacional de Desenvolvimento Cientifico e Tecnologico and the Ministerio da Ciencia, Tecnologia e Inovacao, the Deutsche Forschungsgemeinschaft and the Collaborating Institutions in the Dark Energy Survey. The Collaborating Institutions are Argonne National Laboratory, the University of California at Santa Cruz, the University of Cambridge, Centro de Investigaciones Energeticas, Medioambientales y Tecnologicas-Madrid, the University of Chicago, University College London, the DES-Brazil Consortium, the University of Edinburgh, the Eidgenossische Technische Hochschule (ETH) Zurich, Fermi National Accelerator Laboratory, the University of Illinois at Urbana-Champaign, the Institut de Ciencies de l'Espai (IEEC/CSIC), the Institut de Fisica d'Altes Energies, Lawrence Berkeley National Laboratory, the Ludwig-Maximilians Universitat Munchen and the associated Excellence Cluster Universe, the University of Michigan, the National Optical Astronomy Observatory, the University of Nottingham, the Ohio State University, the University of Pennsylvania, the University of Portsmouth, SLAC National Accelerator Laboratory, Stanford University, the University of Sussex, and Texas A$\&$M University.
\begin{figure*}
\center
\includegraphics[width=1\textwidth]{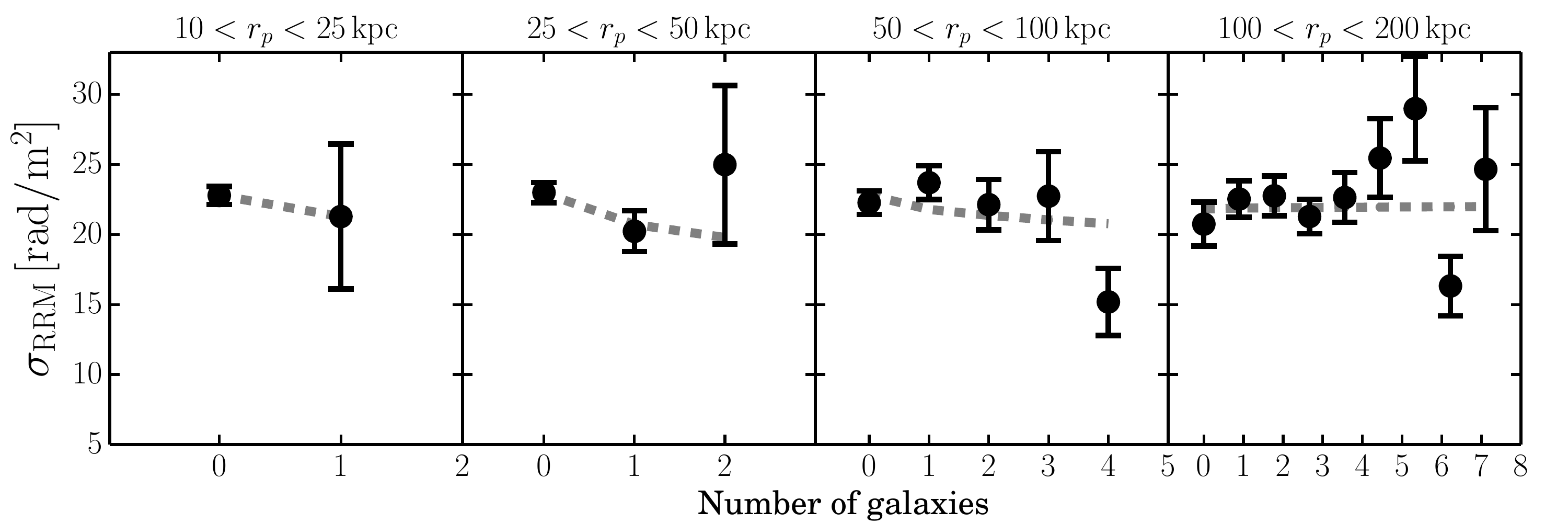}
\caption{Dispersion of RRMs as a function of number of galaxies and impact parameters from 10 kpc to 200 kpc using \citet{Xu2014a} galactic map. The gray dashed lines show the best-fit functions. No correlation between $\sigma_{\rm RRM}$ and $\rm N_{gal}$ is detected at all scales. 
}
\label{fig:Xu}
\end{figure*}

BASS is a key project of the Telescope Access Program (TAP), which has been funded by the National Astronomical Observatories of China, the Chinese Academy of Sciences (the Strategic Priority Research Program "The Emergence of Cosmological Structures" Grant $\#$ XDB09000000), and the Special Fund for Astronomy from the Ministry of Finance. The BASS is also supported by the External Cooperation Program of Chinese Academy of Sciences (Grant $\#$ 114A11KYSB20160057), and Chinese National Natural Science Foundation (Grant $\#$ 11433005).

The Legacy Survey team makes use of data products from the Near-Earth Object Wide-field Infrared Survey Explorer (NEOWISE), which is a project of the Jet Propulsion Laboratory/California Institute of Technology. NEOWISE is funded by the National Aeronautics and Space Administration.

The Legacy Surveys imaging of the DESI footprint is supported by the Director, Office of Science, Office of High Energy Physics of the U.S. Department of Energy under Contract No. DE-AC02-05CH1123, by the National Energy Research Scientific Computing Center, a DOE Office of Science User Facility under the same contract; and by the U.S. National Science Foundation, Division of Astronomical Sciences under Contract No. AST-0950945 to NOAO.

The Photometric Redshifts for the Legacy Surveys (PRLS) catalog used in this paper was produced thanks to funding from the U.S. Department of Energy Office of Science, Office of High Energy Physics via grant DE-SC0007914.
\section{Data availability}
The data underlying this article are available at \url{https://people.ucsc.edu/~tlan3/research/CGM_magnetic_fields/}.

\appendix
\section{Measurements with Xu \& Han (2014) Galactic map}
\label{app:A}
Here we perform our analysis using the RRM dataset from \citet{Xu2014b} with the GRM correction based on the Galactic map of \citet{Xu2014a}.
We use sightlines with $z>1$, $\rm |RRM|<100 \, \rm rad/m^{2}$ and uncertainties of RRMs lower than 20 $\rm rad/m^{2}$, the same selection as applied in the main analysis. The final catalog consists of $\sim 1,300$ systems. Figure~\ref{fig:Xu} shows the results, consistent with the non-detection results as shown in Figure~\ref{fig:sigma_rrm_kpc} based on the Galactic map of \citet{Oppermann2015}. This demonstrates that our results are not sensitive to the foreground Galactic map.

\begin{figure}
\center
\includegraphics[width=0.35\textwidth]{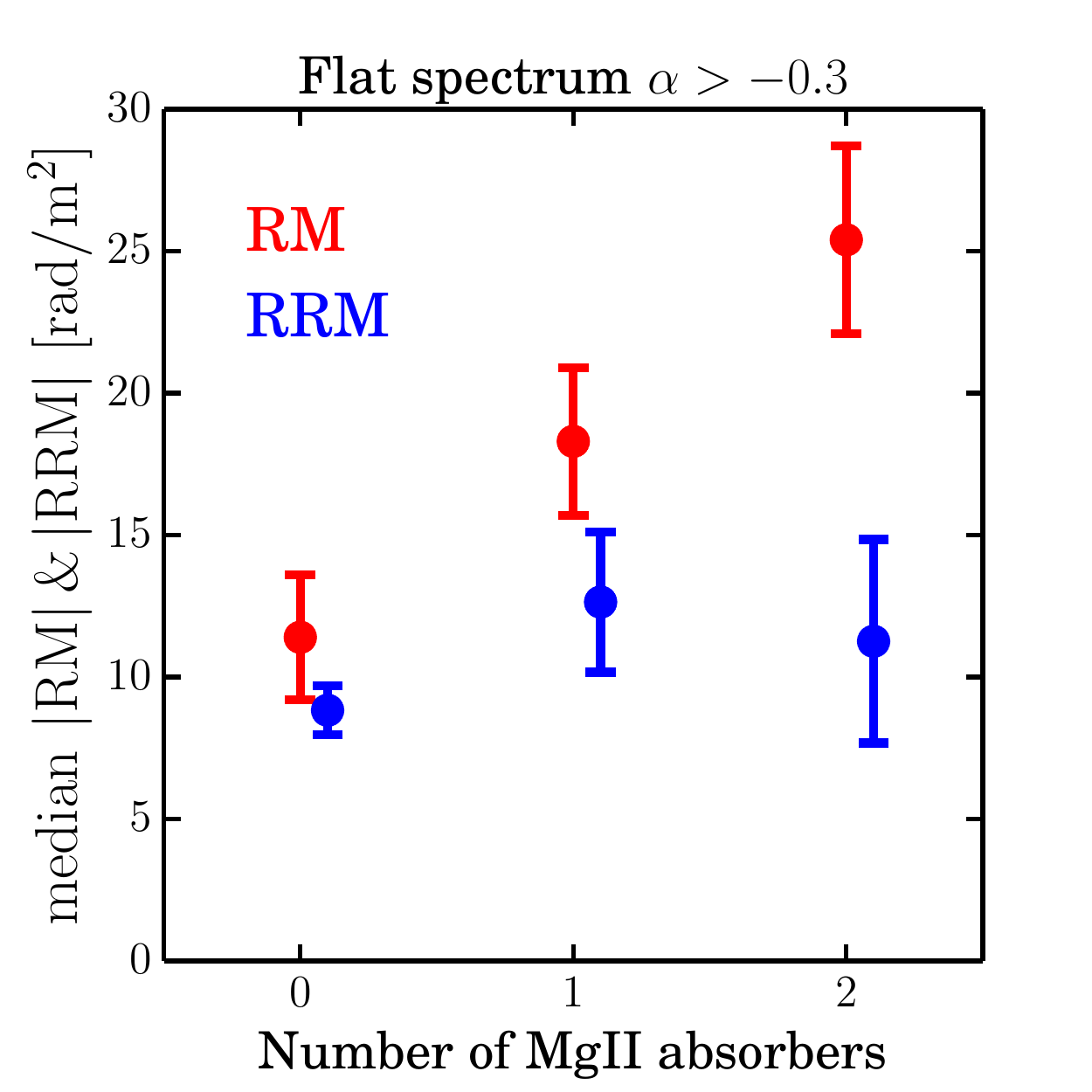}
\caption{Correlation measurements with the \citet{Farnes2014} sample. The median absolute rotation measure as a function of number of MgII absorbers as reported in \citet{Farnes2014} are shown by the red data points. The blue data points show the same measurements with the Galactic contribution, GRM, removed. 
}
\label{fig:Farnes}
\end{figure}

\begin{figure}
\center
\includegraphics[width=0.35\textwidth]{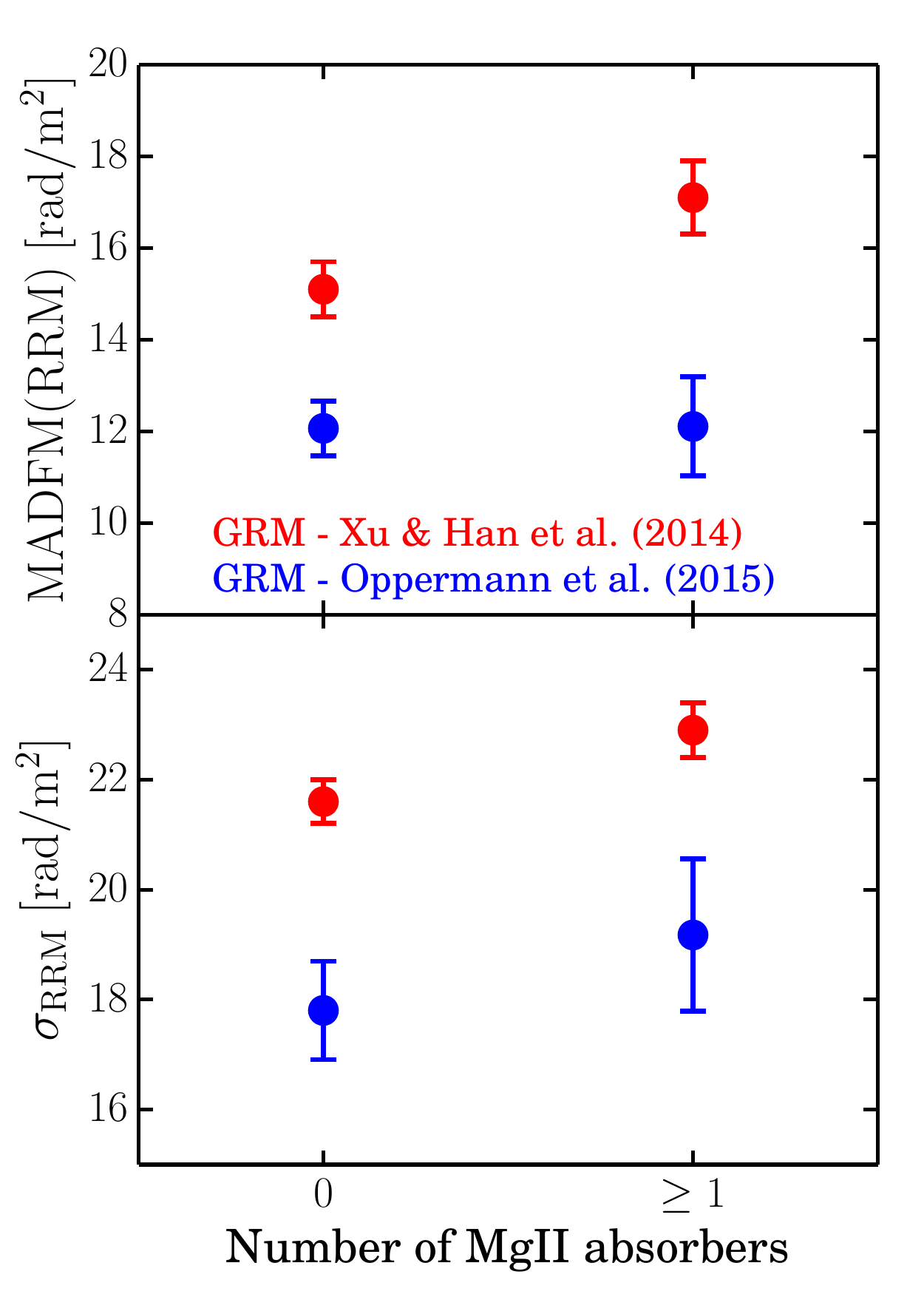}
\caption{Correlation measurements with \citet{Malik2020} sample. The data points in red and blue show the measurements with \citet{Xu2014a} and \citet{Oppermann2015} galactic maps respectively. 
}
\label{fig:Malik}
\end{figure}

\section{Comparison with previous results}
\label{app:B}
By exploring the dispersion of the rotation measures of high redshift radio sources as a function of the number of photometric galaxies in the foreground, we find that no correlation between the two quantifies can be detected with current datasets. 
This non-detection is inconsistent with several results previously reported in the literature. For instance, \citet{Farnes2014} report a $3.5\, \sigma$ detection for the correlation between the presence of MgII absorbers along quasar sightlines and the excess rotation measures. This detection is particularly strong along quasar sightlines with radio spectral index $\alpha>-0.3$. A similar study by \citet{Malik2020} also reports a $4 \sigma$ detection for such correlation. 

To explore the apparent inconsistency between our results and their results, we have reanalyzed the datasets of \citet{Farnes2014} and \citet{Malik2020}. We find that the Galactic foreground contribution may play a dominant
role in their results. In \citet{Farnes2014}, they focused on measuring median $\rm |RM|$ values and argued that the Galactic foreground is not expected to induce the observed correlation. 
However, we notice that the sightlines with absorbers have higher $\rm |GRM|$ in the map of \citet{Oppermann2015}.
Instead of using $\rm |RM|$ values for the measurements, we use $\rm |RRM|$ by adopting the Galactic map from \citet{Oppermann2015} and find that median $\rm |RRM|$ values for quasar sightlines with and without absorbers are consistent with each other. Figure~\ref{fig:Farnes} shows the results. This exercise suggests that ignoring the Galactic foreground might affect the final results.

The choice of Galactic maps also affects the results of \citet{Malik2020}. In \citet{Malik2020}, they make use of the RM galactic map from \citet{Xu2014a} to subtract the Galactic component and estimate (1) the median absolute deviation form the mean, MADFM, and (2) standard deviation of the RRM distribution. Their reported values are shown by the red data points in Figure~\ref{fig:Malik}. The blue data points show the measurements with the RM galactic map from \citet{Oppermann2015}. 
We find that if the \citet{Oppermann2015} map is used, there is no significant correlation between the dispersion of RRMs and the number of absorbers, consistent with our findings.

We note that the dispersion of the measurements is systematically lower when the map of \citet{Oppermann2015} is applied. This is likely due to the fact that the map of \citet{Oppermann2015} has higher angular resolution than the map of \citet{Xu2014a} and therefore it is more sensitive to the small-scale structure of Milky Way magnetic fields.


One of the early results reported by \citet{Bernet2008} showed that the rotation measure from an intervening absorber can be as large as 100 $\rm rad/m^{2}$ in the rest frame and the magnetic field strength $\rm \sim 10 \, \mu G$ (assuming $\rm N_{e}\sim10^{20} \, cm^{-2}$). These values are much higher than our $3\sigma$ upper limits. We note that \citet{Bernet2008} make use of rotation measures obtained at 6 cm, while the catalog we use from \citet{Farnes2014} makes use of rotation measures obtained at 21 cm. One possible mechanism responsible for the inconsistency is the inhomogeneous foreground screens effect, which increases depolarization toward longer wavelengths as discussed in \citet{Bernet2012}. This effect reduces the rotation measure observed in longer wavelengths and therefore gives rise to the discrepancy between the results obtained from 6 cm and 21 cm. On the other hand, the correlation reported in \citet{Bernet2008} is based on a relative small sample with $\sim 70$ sightlines. The correlation signals between the presence of MgII absorbers and the enhance of RM are detected with about 1.7 $\sigma$ significance for sightlines with one MgII absorber and 3.3 $\sigma$ significance for sightlines with two absorbers but from only five sources.
A recent study by \citet{Kim2016} performs a more detailed analysis, using Rotation
Measure (RM) Synthesis, to investigate the Faraday Depth spectra of a sub-sample of \citet{Bernet2008}. 
They find that the possible rotation measure contributed by MgII absorbers is $\sim 20 \rm \ rad/m^{2}$, which brings the measurements closer to the measurements obtained at 21 cm.
To further reconcile the discrepancy between the correlation seen with rotation measures obtained from 6 cm and 21 cm, it will require systematic measurements for sources with radio observations covering a wide range of wavelengths.

{}
\end{CJK*}

\begin{thebibliography}{}
\bibitem[Akahori \& Ryu(2010)]{Akahori2010} Akahori, T., \& Ryu, D.\ 2010, \apj, 723, 476
\bibitem[Amiaux et al.(2012)]{Amiaux2012} Amiaux, J., Scaramella, R., Mellier, Y., et al.\ 2012, \procspie, 8442, 84420Z 
\bibitem[Basu \& Roy(2013)]{Basu2013} Basu, A., \& Roy, S.\ 2013, \mnras, 433, 1675
\bibitem[Beck(2015)]{Beck2015} Beck, R.\ 2015, \aapr, 24, 4
\bibitem[Berlok \& Pfrommer(2019)]{Berlock2019} Berlok, T., \& Pfrommer, C.\ 2019, \mnras, 489, 3368
\bibitem[Bernet et al.(2008)]{Bernet2008} Bernet, M.~L., Miniati, F., Lilly, S.~J., et al.\ 2008, \nat, 454, 302
\bibitem[Bernet et al.(2010)]{Bernet2010} Bernet, M.~L., Miniati, F., \& Lilly, S.~J.\ 2010, \apj, 711, 380
\bibitem[Bernet et al.(2012)]{Bernet2012} Bernet, M.~L., Miniati, F., \& Lilly, S.~J.\ 2012, \apj, 761, 144
\bibitem[Boquien et al.(2019)]{Boquien2019} Boquien, M., Burgarella, D., Roehlly, Y., et al.\ 2019, \aap, 622, A103
\bibitem[Bordoloi et al.(2011)]{Bordoloi2011} Bordoloi, R., Lilly, S.~J., Knobel, C., et al.\ 2011, \apj, 743, 10
\bibitem[Bordoloi et al.(2018)]{Bordoloi2018} Bordoloi, R., Prochaska, J.~X., Tumlinson, J., et al.\ 2018, \apj, 864, 132
\bibitem[Borthakur et al.(2016)]{Borthakur2016} Borthakur, S., Heckman, T., Tumlinson, J., et al.\ 2016, \apj, 833, 259
\bibitem[Bruzual \& Charlot(2003)]{BC2003} Bruzual, G., \& Charlot, S.\ 2003, \mnras, 344, 1000



\bibitem[Carilli \& Rawlings(2004)]{Carilli2004} Carilli, C.~L., \& Rawlings, S.\ 2004, \nar, 48, 979
\bibitem[Chabrier(2003)]{Chabrier2003} Chabrier, G.\ 2003, \pasp, 115, 763
\bibitem[Chandran \& Cowley(1998)]{Chandran1998} Chandran, B.~D.~G., \& Cowley, S.~C.\ 1998, \prl, 80, 3077

\bibitem[Chen et al.(2010)]{Chen2010} Chen, H.-W., Helsby, J.~E., Gauthier, J.-R., et al.\ 2010, \apj, 714, 1521

\bibitem[Dey et al.(2019)]{Dey2019} Dey, A., Schlegel, D.~J., Lang, D., et al.\ 2019, \aj, 157, 168
\bibitem[En{\ss}lin et al.(2009)]{Ensslin2009} En{\ss}lin, T.~A., Frommert, M., \& Kitaura, F.~S.\ 2009, \prd, 80, 105005

\bibitem[Farnes et al.(2014)]{Farnes2014} Farnes, J.~S., O'Sullivan, S.~P., Corrigan, M.~E., et al.\ 2014, \apj, 795, 63
\bibitem[Farnes et al.(2017)]{Farnes2017} Farnes, J.~S., Rudnick, L., Gaensler, B.~M., et al.\ 2017, \apj, 841, 67
\bibitem[Fukugita \& Peebles(2004)]{Fukugita2004} Fukugita, M., \& Peebles, P.~J.~E.\ 2004, \apj, 616, 643

\bibitem[Gaensler et al.(2010)]{POSSUM} Gaensler, B.~M., Landecker, T.~L., Taylor, A.~R., et al.\ 2010, American Astronomical Society Meeting Abstracts \#215 215, 470.13

\bibitem[Govoni et al.(2010)]{Govoni2010} Govoni, F., Dolag, K., Murgia, M., et al.\ 2010, \aap, 522, A105


\bibitem[Hammond et al.(2012)]{Hammond2012} Hammond, A.~M., Robishaw, T., \& Gaensler, B.~M.\ 2012, arXiv e-prints, arXiv:1209.1438
\bibitem[Han(2017)]{Han2017} Han, J.~L.\ 2017, \araa, 55, 111
\bibitem[Heckman et al.(2017)]{Heckman2017} Heckman, T., Borthakur, S., Wild, V., et al.\ 2017, \apj, 846, 151
\bibitem[Hopkins et al.(2020)]{Hopkins2020} Hopkins, P.~F., Chan, T.~K., Ji, S., et al.\ 2020, arXiv e-prints, arXiv:2002.02462
\bibitem[Hummels et al.(2019)]{Hummels2019} Hummels, C.~B., Smith, B.~D., Hopkins, P.~F., et al.\ 2019, \apj, 882, 156
\bibitem[Ivezi{\'c} et al.(2019)]{Ivezic2019} Ivezi{\'c}, {\v{Z}}., Kahn, S.~M., Tyson, J.~A., et al.\ 2019, \apj, 873, 111


\bibitem[Ji et al.(2018)]{Ji2018} Ji, S., Oh, S.~P., \& McCourt, M.\ 2018, \mnras, 476, 852
\bibitem[Joshi \& Chand(2013)]{Joshi2013} Joshi, R., \& Chand, H.\ 2013, \mnras, 434, 3566

\bibitem[Kacprzak et al.(2012)]{Kac2012} Kacprzak, G.~G., Churchill, C.~W., \& Nielsen, N.~M.\ 2012, \apjl, 760, L7
\bibitem[Kim et al.(2016)]{Kim2016} Kim, K.~S., Lilly, S.~J., Miniati, F., et al.\ 2016, \apj, 829, 133

\bibitem[Klein \& Fletcher(2015)]{Klein2015} Klein, U., \& Fletcher, A.\ 2015, Galactic and Intergalactic Magnetic Fields

\bibitem[Kronberg \& Perry(1982)]{Kronberg1982} Kronberg, P.~P., \& Perry, J.~J.\ 1982, \apj, 263, 518
\bibitem[Kronberg et al.(2008)]{Kronberg2008} Kronberg, P.~P., Bernet, M.~L., Miniati, F., et al.\ 2008, \apj, 676, 70


\bibitem[Lan et al.(2014)]{Lan2014} Lan, T.-W., M{\'e}nard, B., \& Zhu, G.\ 2014, \apj, 795, 31
\bibitem[Lan \& Fukugita(2017)]{Lan2017} Lan, T.-W., \& Fukugita, M.\ 2017, \apj, 850, 156

\bibitem[Lan \& Mo(2018)]{Lan2018} Lan, T.-W., \& Mo, H.\ 2018, \apj, 866, 36
\bibitem[Lan \& Mo(2019)]{LanMo2019} Lan, T.-W., \& Mo, H.\ 2019, \mnras, 486, 608
\bibitem[Lan(2019)]{Lan2019} Lan, T.-W.\ 2019, arXiv e-prints, arXiv:1911.01271
\bibitem[Lang et al.(2016)]{Lang2016} Lang, D., Hogg, D.~W., \& Mykytyn, D.\ 2016, The Tractor: Probabilistic astronomical source detection and measurement, ascl:1604.008
\bibitem[Levi et al.(2013)]{Levi2013} Levi, M., Bebek, C., Beers, T., et al.\ 2013, arXiv:1308.0847
\bibitem[Li et al.(2018)]{Li2018} Li, J.-T., Bregman, J.~N., Wang, Q.~D., et al.\ 2018, \apjl, 855, L24
\bibitem[Liang \& Remming(2020)]{Liang2020} Liang, C.~J., \& Remming, I.\ 2020, \mnras, 491, 5056
\bibitem[Lim et al.(2020)]{Lim2020} Lim, S.~H., Mo, H.~J., Wang, H., et al.\ 2020, \apj, 889, 48


\bibitem[Malik et al.(2020)]{Malik2020} Malik, S., Chand, H., \& Seshadri, T.~R.\ 2020, \apj, 890, 132
\bibitem[Marinacci et al.(2018)]{Marinacci2018} Marinacci, F., Vogelsberger, M., Pakmor, R., et al.\ 2018, \mnras, 480, 5113
\bibitem[Mao et al.(2017)]{Mao2017} Mao, S.~A., Carilli, C., Gaensler, B.~M., et al.\ 2017, Nature Astronomy, 1, 621

\bibitem[McCourt et al.(2015)]{McCourt2015} McCourt, M., O'Leary, R.~M., Madigan, A.-M., et al.\ 2015, \mnras, 449, 2

\bibitem[Nelson(1973)]{Nelson1973} Nelson, A.~H.\ 1973, \pasj, 25, 489
\bibitem[Nelson et al.(2018)]{Nelson2018} Nelson, D., Kauffmann, G., Pillepich, A., et al.\ 2018, \mnras, 477, 450
\bibitem[Nelson et al.(2020)]{Nelson2020} Nelson, D., Sharma, P., Pillepich, A., et al.\ 2020, arXiv e-prints, arXiv:2005.09654
\bibitem[Neronov et al.(2013)]{Neronov2013} Neronov, A., Semikoz, D., \& Banafsheh, M.\ 2013, arXiv e-prints, arXiv:1305.1450
\bibitem[Nielsen et al.(2013)]{Nielsen2013} Nielsen, N.~M., Churchill, C.~W., \& Kacprzak, G.~G.\ 2013, \apj, 776, 115

\bibitem[Oppermann et al.(2011)]{Oppermann2011} Oppermann, N., Robbers, G., \& En{\ss}lin, T.~A.\ 2011, \pre, 84, 041118

\bibitem[Oppermann et al.(2012)]{Oppermann2012} Oppermann, N., Junklewitz, H., Robbers, G., et al.\ 2012, \aap, 542, A93

\bibitem[Oppermann et al.(2015)]{Oppermann2015} Oppermann, N., Junklewitz, H., Greiner, M., et al.\ 2015, \aap, 575, A118
\bibitem[O'Sullivan et al.(2020)]{Sullivan2020} O'Sullivan, S.~P., Br{\"u}ggen, M., Vazza, F., et al.\ 2020, arXiv e-prints, arXiv:2002.06924
\bibitem[Oren \& Wolfe(1995)]{Oren1995} Oren, A.~L., \& Wolfe, A.~M.\ 1995, \apj, 445, 624

\bibitem[Pakmor \& Springel(2013)]{Pakmor2013} Pakmor, R., \& Springel, V.\ 2013, \mnras, 432, 176

\bibitem[Pakmor et al.(2019)]{Pakmor2019} Pakmor, R., van de Voort, F., Bieri, R., et al.\ 2019, arXiv e-prints, arXiv:1911.11163
\bibitem[Ravi et al.(2019)]{Ravi2019} Ravi, V., Battaglia, N., Burke-Spolaor, S., et al.\ 2019, \baas, 51, 420

\bibitem[P{\'e}roux et al.(2018)]{Peroux2018} P{\'e}roux, C., Rahmani, H., Arrigoni Battaia, F., et al.\ 2018, \mnras, 479, L50


\bibitem[Prochaska et al.(2017)]{Prochaska2017} Prochaska, J.~X., Werk, J.~K., Worseck, G., et al.\ 2017, \apj, 837, 169

\bibitem[Prochaska et al.(2019)]{Prochaska2019} Prochaska, J.~X., Macquart, J.-P., McQuinn, M., et al.\ 2019, Science, 366, 231

\bibitem[Rubin et al.(2018)]{Rubin2018} Rubin, K.~H.~R., Diamond-Stanic, A.~M., Coil, A.~L., et al.\ 2018, \apj, 868, 142

\bibitem[Ryu et al.(2008)]{Ryu2008} Ryu, D., Kang, H., Cho, J., et al.\ 2008, Science, 320, 909

\bibitem[Salem et al.(2016)]{Salem2016} Salem, M., Bryan, G.~L., \& Corlies, L.\ 2016, \mnras, 456, 582
\bibitem[Schnitzeler(2010)]{Schnitzeler2010} Schnitzeler, D.~H.~F.~M.\ 2010, \mnras, 409, L99
\bibitem[Schroetter et al.(2019)]{Schroetter2019} Schroetter, I., Bouch{\'e}, N.~F., Zabl, J., et al.\ 2019, \mnras, 490, 4368

\bibitem[Stocke et al.(2013)]{Stocke2013} Stocke, J.~T., Keeney, B.~A., Danforth, C.~W., et al.\ 2013, \apj, 763, 148
\bibitem[Stuardi et al.(2020)]{Stuardi2020} Stuardi, C., O'Sullivan, S.~P., Bonafede, A., et al.\ 2020, arXiv e-prints, arXiv:2004.05169
\bibitem[Taylor et al.(2009)]{Taylor2009} Taylor, A.~R., Stil, J.~M., \& Sunstrum, C.\ 2009, \apj, 702, 1230
\bibitem[Tumlinson et al.(2013)]{Tumlinson2013} Tumlinson, J., Thom, C., Werk, J.~K., et al.\ 2013, \apj, 777, 59
\bibitem[Tumlinson et al.(2017)]{Tumlinson2017} Tumlinson, J., Peeples, M.~S., \& Werk, J.~K.\ 2017, \araa, 55, 389

\bibitem[van de Voort et al.(2019)]{Voort2019} van de Voort, F., Springel, V., Mandelker, N., et al.\ 2019, \mnras, 482, L85
\bibitem[van Haarlem et al.(2013)]{Haarlem2013} van Haarlem, M.~P., Wise, M.~W., Gunst, A.~W., et al.\ 2013, \aap, 556, A2
\bibitem[Vazza et al.(2018)]{Vazza2018} Vazza, F., Br{\"u}ggen, M., Hinz, P.~M., et al.\ 2018, \mnras, 480, 3907
\bibitem[Vernstrom et al.(2019)]{Vernstrom2019} Vernstrom, T., Gaensler, B.~M., Rudnick, L., et al.\ 2019, \apj, 878, 92

\bibitem[Welter et al.(1984)]{Welter1984} Welter, G.~L., Perry, J.~J., \& Kronberg, P.~P.\ 1984, \apj, 279, 19

\bibitem[Werk et al.(2014)]{Werk2014} Werk, J.~K., Prochaska, J.~X., Tumlinson, J., et al.\ 2014, \apj, 792, 8
\bibitem[Wolfe et al.(1992)]{Wolfe1992} Wolfe, A.~M., Lanzetta, K.~M., \& Oren, A.~L.\ 1992, \apj, 388, 17

\bibitem[Xu \& Han(2014a)]{Xu2014a} Xu, J., \& Han, J.~L.\ 2014, \mnras, 442, 3329
\bibitem[Xu \& Han(2014b)]{Xu2014b} Xu, J., \& Han, J.-L.\ 2014, Research in Astronomy and Astrophysics, 14, 942-958

\bibitem[Zhou et al.(2020)]{Zhou2020} Zhou, R., Newman, J.~A., Mao, Y.-Y., et al.\ 2020, arXiv e-prints, arXiv:2001.06018

\bibitem[Zhu \& M{\'e}nard(2013)]{Zhu2013b} Zhu, G., \& M{\'e}nard, B.\ 2013, \apj, 773, 16
\bibitem[Zou et al.(2017)]{Zou2017} Zou, H., Zhou, X., Fan, X., et al.\ 
2017, \pasp, 129, 064101



\end{thebibliography}
\end{document}